\documentclass[epj]{svjour}

\usepackage{amssymb}
\usepackage{amsmath}
\usepackage{graphicx}
\usepackage{bm}
\usepackage{cuted}
\usepackage{hyperref}

\begin{document}

\title{Magnets in an electric field: hidden forces and momentum conservation}
\author{Francis Redfern}
\institute{Texarkana College, Texarkana, Texas 75599}
\abstract{
In 1967 Shockley and James addressed the situation of a magnet in an electric
field. The magnet is at rest and contains electromagnetic momentum, but there
was no obvious mechanical momentum to balance this for momentum conservation.
They concluded that some sort of mechanical momentum, which they called
``hidden momentum'', was contained in the magnet and ascribed this momentum to
relativistic effects, a contention that was apparently confirmed by Coleman and
Van Vleck. Since then, a magnetic dipole in an electric field has been
considered to have this new form of momentum, but this view ignores the
electromagnetic forces that arise when an electric field is applied to a magnet
or a magnet is formed in an electric field. The electromagnetic forces result
in the magnet gaining electromagnetic momentum and an equal and opposite amount
of mechanical momentum so that it is moving in its original rest frame. This
moving reference frame is erroneously taken to be the rest frame in studies
that purport to show hidden momentum. Here I examine the analysis of Shockley
and James and of Coleman and Van Vleck and consider a model of a magnetic
dipole formed in a uniform electric field. These calculations show no hidden
momentum.}

\maketitle

\begin{strip}
The final version is available at Springer via
\href{http://dx.doi.org/10.1140/epjd/e2017-70263-3}{Redfern, F. Eur. Phys. J. D (2017) 71: 163. https://doi.org/10.1140/epjd/e2017-70263-3}
\end{strip}

\section{Introduction}

In 1891 J.J. Thompson pointed out an apparent paradox \cite{JJ91}. It appeared
as if electromagnetic systems at rest could contain non-zero electromagnetic
momentum. Later \cite{JJ04}, he calculated the (Lorentz) electromagnetic
momentum of an at-rest system consisting of a point charge in the vicinity of an
Amperian dipole (for example, a loop of current) to be, in SI units,
$\epsilon_o \bm{E}\times\bm{B}V$, where $\bm{E}$ is the electric field at the
magnetic dipole, $\bm{B}$ is the uniform magnetic field inside the solenoid he
used to approximate the Amperian dipole, and $V$ is the solenoid's volume.
However, it does not appear he ever postulated a mechanical momentum, equal and
opposite that of the electromagnetic momentum, presumably necessary to preserve
momentum conservation \cite{McDonald}.

In their 1967 paper, Shockley and James \cite{Shock} introduced the term
``hidden momentum'' to describe this ``hitherto disregarded momentum'' they
thought necessary to conserve linear momentum in their charge and magnet
model. In this model there are two small spheres of opposite charge in the
vicinity of a magnet consisting of counter-rotating oppositely charged disks
contained within a ``pill box''. They were puzzled by the
lack of mechanical momentum to balance the electromagnetic momentum in this
at-rest system. Then when the magnet is demagnetized by bringing the rotating
disks slowly to rest, the force they calculated acting on the charged spheres
was not accompanied by an obvious equal and opposite force on the magnet
necessary to conserve linear momentum. They proposed the existence of mechanical
hidden momentum in the rotating disks that balanced the electromagnetic momentum
before they were brought to rest. The release of this momentum as the disks
slowed, they conjectured, would result in a force acting on the magnet. They
speculated that relativistic effects were involved in this momentum.

A year later Coleman and Van Vleck \cite{Coleman} published a detailed analysis
on the question of a point charge in the vicinity of a magnet to see if the
conjecture of Shockley and James was correct. They based their analysis on a
Lagrangian for electromagnetic systems derived by Darwin \cite{Darwin}, and
concluded that the demagnetizing
magnet would experience an equal and opposite impulse to that
experienced by the point charge and equal in magnitude to that found by Shockley
and James. Hence, they conclude momentum is conserved in this system and the
hidden momentum is due to relativistic effects as proposed by Shockley and
James. Unlike the magnet of Shockley and James, that of Coleman and Van Vleck
was not model dependent, though they alluded to a specific model in footnote 9
of their paper.

In this paper, rather than beginning with a point charge-magnet system already
intact, I apply the electric field to the magnet of Shockley and James by
bringing in the charges originally very distant from the pill box magnet. There
can be no question that there is zero momentum, electromagnetic and
mechanical, as well as zero interaction energy, for this starting point. I
show that upon assembly of the system (hereafter called the SJ model) Lorentz
forces arise such that, if you want to keep this system at rest, you need an
external source of mechanical force to counteract the electromagnetic forces.

Hence, to the charges and magnet in this reference frame you must add an
external agent -- let us say initially containing zero momentum -- that can
exert forces on the charges and magnet as the SJ model is assembled. You end up
with the external agent having a mechanical momentum equal to the mechanical
momentum which the SJ model would acquire were it not constrained to remain at
rest, equal and opposite to the electromagnetic momentum in the
system.

Thus if the SJ model contains hidden momentum equal and opposite to its
electromagnetic momentum after being held stationary, you get the following
paradox. The momenta of all components of the model in the original rest frame
were zero, as was the momentum of the external agent, but now, in the same rest
frame, the sum of the mechanical momentum of the external agent, the
electromagnetic momentum of the SJ model, and the hidden mechanical momentum is
no longer zero.

Next I revisit the paper of Coleman and Van Vleck and explicitly calculate the
sums in the Darwin Lagrangian of the point charge-magnet system for a model
where the current consists of non-interacting
charges that flow along a circular frictionless track. (This is the model they
alluded to in footnote 9.) I will show that the hidden momentum of the model
only appears to be present because the calculations are, in effect, carried out
in the rest frame of the track rather than in the center of momentum frame of
the system. The effect of this is that the model turns out to be very contrived
and does not correspond to the application of an electric field to a uniform
current loop of non-interacting particles.

In my final calculation I consider the formation of a magnetic dipole inside a
spherical shell with a dipolar electric charge distribution. This distribution
produces a uniform electric field inside the shell and a dipolar electric field
on the outside. The mechanical momentum acquired by the shell when the magnetic
dipole is formed is equal and opposite to the momentum in the resulting
electromagnetic field. No hidden momentum is present. (All accelerations in the
calculations of this paper are assumed to be small enough that radiation
effects may be ignored.) After this calculation, due to suggestions by a
referee, I have added comments on a number of papers, some very recent.

An argument often made for the presence of hidden momentum is that the
center-of-energy theorem \cite{Griff} of relativity requires its presence.
Some physicists with whom I have had private communications have claimed that
it is not necessary to consider the assembly of the charge-magnet system; that
is, applying an electric field to a magnet or creating a magnet in an electric
field. They feel you can treat the system with this theorem without worrying
about how the system came to be.

This is undoubtedly true for many systems. In the case of this sort of system,
however, its formation imparts mechanical momentum to the system (equal and
opposite to the electromagnetic momentum it receives), which will change
its inertial reference frame unless held stationary. If held stationary, energy
and momentum are exchanged with the system's environment due to the necessity
of countering the Lorentz forces when either an
electric field is applied to a magnet or a magnet is formed in an electric
field. To treat the restrained system as if were isolated in the universe, so
to speak, is invalid in my view. Somewhere in this ``universe'' is the momentum
that was exchanged, so to say the system must have zero momentum to satisfy the
center-of-energy theorem ignores this exchange.

\section{The assembly of the model of Shockley and James and the resulting
momentum}\label{model of SJ}

To calculate the electromagnetic field momentum in the Lorentz formulation, you
evaluate the volume integral of the Lorentz electromagnetic field momentum
density. This integral can be written in SI units as follows
\begin{equation}
\label{em momentum}
P_{em} = \epsilon_o\int_V(\bm{E}\times\bm{B})dV,
\end{equation}
where $\bm{E}$ is the electric field and $\bm{B}$ the magnetic field. When this
momentum is nonzero and the system containing this momentum is at rest, it is
generally thought \cite{Griff,Babson} the electromagnetic momentum must be
balanced by
an equal amount of hidden momentum somehow mechanically present in the system.
The problem with the models in which hidden momentum is inferred is the lack of
appreciation of the effect of the Lorentz forces involved in assembling the
models (namely, applying an electric field to a magnet or forming a magnet in
an electric field). When the momentum imparted to the model
in its assembly is taken into account, the total momentum, electromagnetic plus
mechanical, is conserved without the need to postulate a hidden form.

\begin{figure}[ht!]
\label{SJ model}
\centering
\includegraphics[width=4.0in]{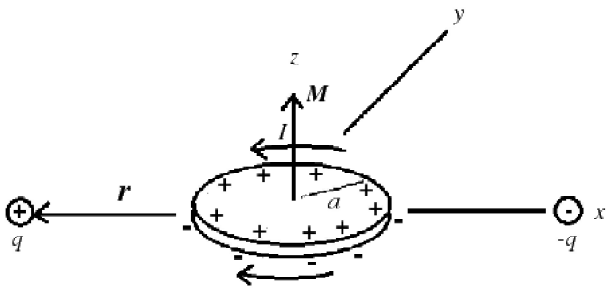}
\caption{A schematic of the model investigated by Shockley and James. See the
text for a description.}
\end{figure}

To illustrate this, I will assemble the model of Shockley and James
\cite{Shock}. They considered a model (Fig. 1) consisting of two ``plastic''
coaxial disks at the origin of a coordinate system, their areas perpendicular
to the $z$ direction with opposite charge distributions on their rims and
rotating in opposite directions such that they produce a current $I$. This
establishes a magnetic moment $\mathcal{M} = IA$ at the origin pointing in the
positive $z$ direction where $A$ is the area of a disk.

The rotating disks are
housed in a pill box with arms extending outward a distance $r$ along the
positive and negative $x$ axis. At the end of these arms are two small
stationary spheres containing equal and opposite charges $\pm q$, the positive
charge on the sphere in the negative $x$ direction and the
negative charge in the positive direction. They imagine there is a brake
applying torque to the disks, acting to bring them
slowly to rest such that little radiation occurs. The spheres holding the
charges are massive enough that they hardly move when the
changing magnetic field produces a force on the spheres,
resulting in negligible radiation.

For this situation they calculate that both spheres receive an impulse of
momentum in the negative $y$ direction with the total impulse being, in SI
units,
\begin{equation}
\label{S&J impulse}
\Delta\bm{P}_{SJ} = -\frac{\mu_o qIA}{2\pi r^2}\bm{\hat{j}} = 2\mu_o\epsilon_o\bm{E}\times\bm{\mathcal{M}},
\end{equation}
where $\bm{E}$ is the electric field due to one charge at the location of the
magnetic moment
$\bm{{\mathcal{M}}} = IA\bm{\hat{k}}$. They argue that since there is no
electromagnetic force on the magnet, there must be hidden mechanical momentum
in the disks, equal and opposite to that obtained by the charges, which is
released to the disks when they are brought to a stop. They also argue that
this hidden momentum is necessary for the momentum of the stationary system to
be zero before the brake is applied to the disks. However, in this model they
neglect what occurs during the application of the electric field (or creating
the magnet in an existing field).

Imagine assembling the SJ model so that mechanical external forces prevent it
from acquiring mechanical momentum. You bring the two charged spheres from a
great distance along the positive and negative $x$ axis to a distance $r$ from
the rotating disks. The external forces needed to move the spheres are equal
and opposite and impart a net momentum (linear or angular) neither to the SJ
model nor to the agent external to the model responsible for the origin of these
forces. However, the charges will experience Lorentz forces in the positive $y$
direction due to traveling through the magnetic field of the disks.

Additionally, the displacement current present at the disks due to
the motion of the charges will result in Lorentz forces acting on the current
$I$, also in the positive $y$ direction. In order to keep the model at rest, the
external agent must apply, in addition to the forces moving the spheres toward
the disks, mechanical forces on the charge-disk system
in the negative $y$ direction that cancel the Lorentz forces. The external
agent will experience an impulse equal and opposite to that in
Eq. (\ref{S&J impulse}) as will be shown below.

Each charge in the SJ model interacting with the magnetic moment will, as is
well known \cite{Jackson2}, contribute an amount of momentum
($\epsilon_o\mu_o \bm{E}\times\bm{\mathcal{M}}$) stored in the electromagnetic
field. The charges approach the disks along the $x$ axis, in the ``equator'' of
the magnetic moment, so the magnetic field at their positions a distance $r$
from the disks is
\begin{equation}
\label{B at charges}
\bm{B} = -\frac{\mu_o IA}{4\pi r^3}\bm{\hat{k}}.
\end{equation}
The Lorentz force experienced by each charge is of the same magnitude and
direction. The total Lorentz force on the charges is
\begin{equation}
\label{F at charges}
\bm{F} = 2qv\bm{\hat{i}}\times (-\frac{\mu_o IA}{4\pi r^3}\bm{\hat{k}}) = \frac{\mu_o qIAv}{2\pi r^3}\bm{\hat{j}},
\end{equation}
where $v$ is the speed of the charges. When you integrate the force over time
you get the impulse delivered to the charges,
\begin{equation}
\label{impulse to charges}
\Delta\bm{P} = \frac{\mu_o qIA}{4\pi r^2}\bm{\hat{j}} = -\mu_o\epsilon_o\bm{E}\times\bm{\mathcal{M}},
\end{equation}
a result also obtained by Boyer \cite{Boyer3} for different model.

\begin{figure}[ht!]
\label{displacement current field}
\centering
\includegraphics[width=4.5in]{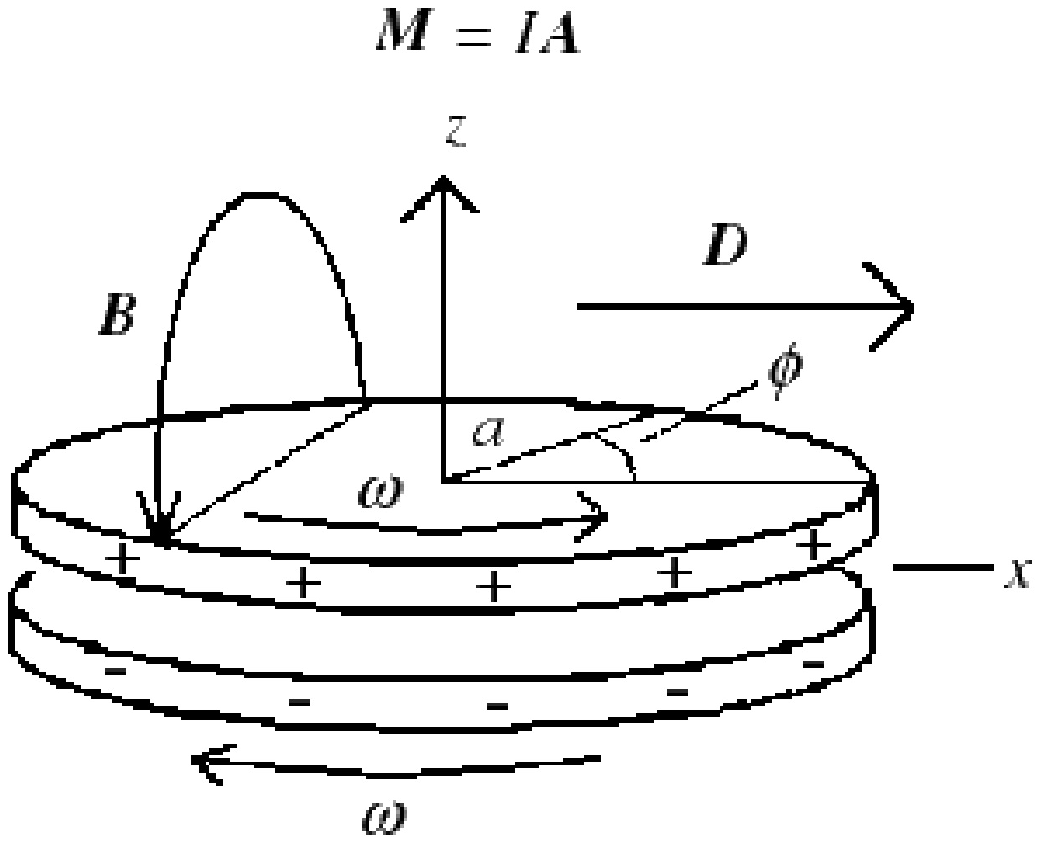}
\caption{The geometry involved in computing the impulse applied to the disks
from the effect of the increasing electric field due to the charges moving
toward them from a distance. $\omega$ is the angular speed of the disks.}
\end{figure}

The next task is to compute the impulse delivered to the disks through the
action of the displacement current. (Refer to Fig. 2.) Assuming the disks are
sufficiently small, the displacement at their position due to both charges is
\begin{equation}
\label{displacement}
\bm{D} = \frac{1}{2\pi}\frac{q}{\gamma^2 r^2}\bm{\hat{i}} \approx \frac{1}{2\pi}\frac{q}{r^2}\bm{\hat{i}},
\end{equation}
for the slow-motion case where $\gamma \approx 1$.
The displacement current responsible for the magnetic field
at the edges of the disks at a given point along the $x$ axis will
depend on the cross-sectional area defined by a
circle with a diameter equal to the distance between the edges of the disks
parallel to $y$. (See Fig. 2.) This cross-sectional area is $Asin^2\phi = 
\pi a^2 sin^2\phi$, where $a$ is the radius of the disk and $\phi$ is the usual
azimuth angle of a spherical coordinate system. The displacement current as a
function of $\phi$ is therefore
\begin{equation}
\label{displacement current}
I_D = \frac{d\bm{D}}{dt}\cdot\bm{\hat{i}}\pi a^2 sin^2\phi = \frac{qva^2}{r^3}sin^2\phi,
\end{equation}
where $\bm{v} = (-dr/dt)\bm{\hat{i}}$ when you take the time
derivative of $\bm{D}$. From the integral form of Ampere's law, you find the
magnetic field at the rims of the disks as a function of $\phi$,
\begin{equation}
\label{B on edges}
\bm{B}_D = \frac{\mu_o I_D}{2\pi asin\phi}\bm{\hat{k}} = \frac{\mu_o qav}{2\pi r^3}sin\phi\bm{\hat{k}}.
\end{equation}

To find the Lorentz force on the disks, you perform the following integration.
\begin{equation}
\label{F on edges}
\bm{F}_D = I\oint d\bm{l}\times\bm{B}_D = I\oint ad\phi\bm{\hat{\phi}}\times\bm{B}_D = \frac{\mu_o qIAv}{2\pi r^3}\bm{\hat{j}},
\end{equation}
where $\bm{\hat{\phi}} = -sin\phi\bm{\hat{i}} + cos\phi\bm{\hat{j}}$. When you
integrate this over time to get the impulse, you find
\begin{equation}
\label{impulse to disks}
\Delta\bm{P}_D = \frac{\mu_o q IA}{4\pi r^2}\bm{\hat{j}} = -\mu_o\epsilon_o\bm{E}\times\bm{\mathcal{M}}.
\end{equation}
Note that this is equal to the impulse applied to the charges, Eq. (\ref{impulse
to charges}), both in magnitude and direction. (This result was also previously
obtained by Boyer using a different approach \cite{Boyer3}.)

To get the total impulse applied to the system by its assembly, you add
Eq. (\ref{impulse to charges}) to Eq.(\ref{impulse to disks}). The result is
equal and opposite that of Eq. (\ref{S&J impulse}), meaning if the SJ model
were held stationary by an external agent, the external agent would acquire the
negative of the momentum given in Eq. (\ref{S&J impulse}).
When the disks are brought to rest, the momentum that was stored in
the electromagnetic field is converted to an impulse applied to the model,
given by Eq. (\ref{S&J impulse}), but this time only to the charges, not to the
disks.

The linear momentum stored in the electromagnetic field is thus converted
into mechanical momentum equal and opposite to that possessed by the external
agent. The total momentum, linear and angular, of the SJ system plus external
agent remains zero throughout this process, and no hidden momentum due to
stresses in the disks as supposed by Shockley and James is necessary
to conserve linear momentum. Rather the ``hidden momentum'' is contained in the
external agent.

\section{A model-dependent examination of the Coleman-Van Vleck calculation}\label{darwin lagrangian}

The Darwin Lagrangian used by Coleman and Van Vleck \cite{Coleman} needs to be
slightly reformulated to be usable in calculations of the model I examine.
Referring to the system of Fig. 3, where there is a single charge in the
vicinity of a magnet, the Lagrangian becomes, in SI units (symbols defined
below),
\begin{eqnarray}
\label{Darwin L}
L &=& \frac{1}{2}mv^2 + \frac{1}{8c^2}mv^4 + \frac{1}{2}{M}V^2 + \frac{1}{8c^2}MV^4 \\ \nonumber
&+& \frac{1}{2}\sum_i m_i v_i^2 + \frac{1}{8c^2}\sum_i m_i v_i^4 \\ \nonumber
&-& \frac{q}{4\pi\epsilon_o}\sum_i\frac{q_i}{|\bm{r} - \bm{a}'_i|} - \frac{1}{8\pi\epsilon_o}\sum_{i\neq j}\frac{q_i q_j}{|\bm{a}'_i - \bm{a}'_j|} \\ \nonumber
&+& \frac{q}{8\pi\epsilon_o c^2}\sum_i\frac{q_i}{|\bm{r} - \bm{a}'_i|} \left[\bm{v}\cdot\bm{v}'_i + \frac{(\bm{v}\cdot(\bm{r} - \bm{a}'_i))(\bm{v}'_i\cdot(\bm{r} - \bm{a}'_i))}{|\bm{r} - \bm{a}'_i|^2}\right] \\ \nonumber
&+& \frac{1}{8\pi\epsilon_o c^2}\sum_{i\neq j}\frac{q_i q_j}{|\bm{a}'_i - \bm{a}'_j|} \left[\bm{v}'_i\cdot\bm{v}'_j + \frac{(\bm{v}'_i\cdot(\bm{a}'_i - \bm{a}'_j))(\bm{v'}_j\cdot(\bm{a}'_i - \bm{a}'_j))}{|\bm{a}'_i - \bm{a}'_j|^2}\right].
\end{eqnarray}

\begin{figure}[ht!]
\label{CVV model}
\centering
\includegraphics[width=3in]{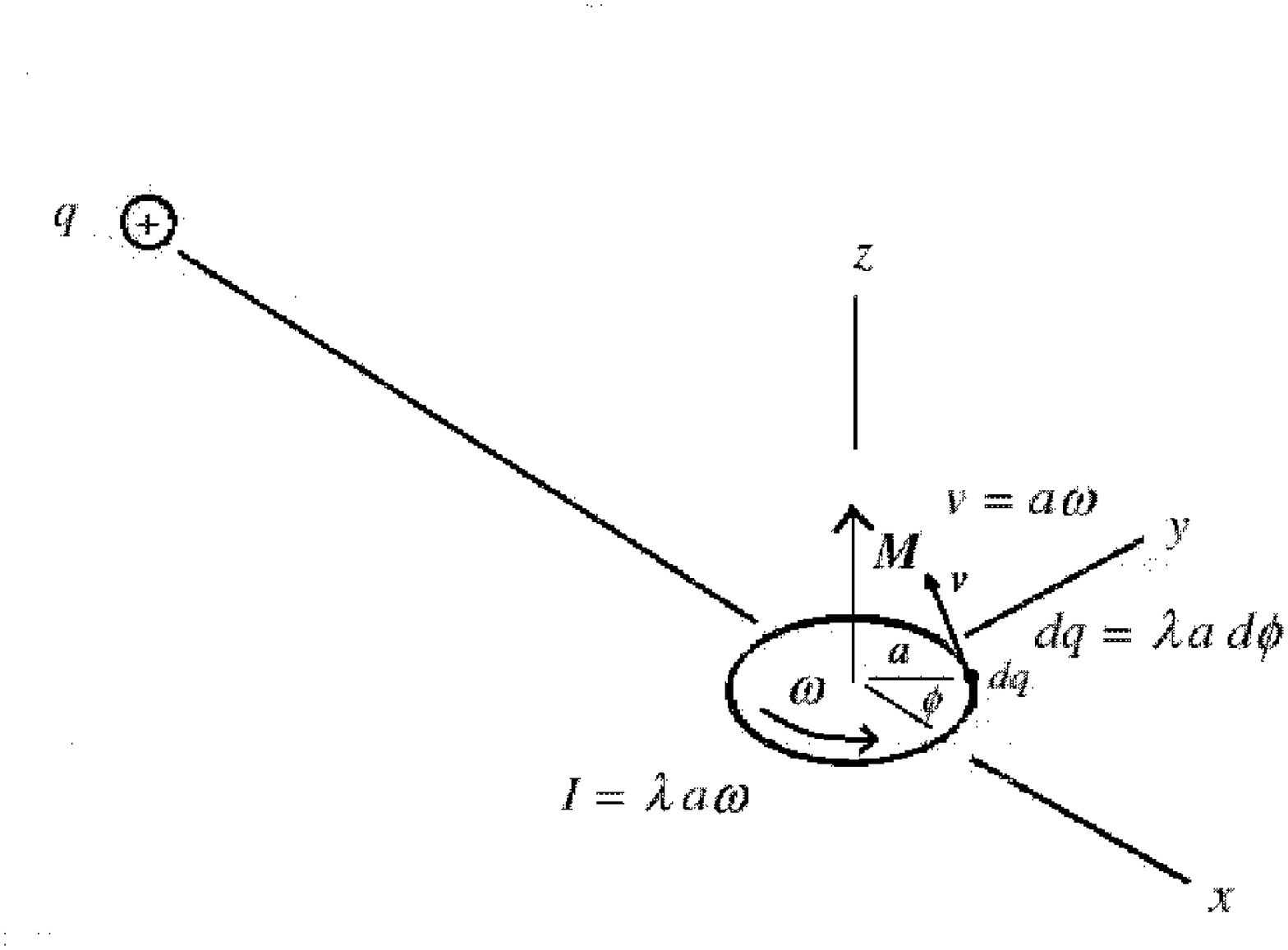}
\caption{The charge-magnet geometry used for the Darwin Lagrangian calculations.}
\end{figure}

The model addressed in this section assumes the current producing the magnetism
consists of a frictionless circular tube with the electrical properties of the
vacuum, a radius equal to that of the magnet, and containing non-interacting
positive charges moving in the positive sense in the $x$-$y$ plane.
This positive charge is neutralized by a coaxial ring of negative charge of the
same radius that is stationary. The origin of the coordinate system is the
position of the center of the magnet. The third and fourth summations in the
above Lagrangian can be considered zero if, as is assumed by both Shockley and
James and by Coleman and Van Vleck, the magnet is neutral and electrically
unpolarized. In this equation (see also Fig. 3) the quantities and variables
are defined as follows.
\begin{itemize}
\item $c =$ the speed of light,
\item $m =$ the mass of the point charge $q$,
\item $\bm{v} =$ the velocity of the point charge,
\item $q =$ the positive point charge,
\item $M =$ the mass of the magnet,
\item $\bm{V} =$ the velocity of the center of mass of the magnet,
\item $\bm{r} =$ the position of the point charge (in the $-x$ direction) from the origin,
\item $q_i =$ a (tiny) charge of a magnet particle,
\item $m_i =$ the mass of $q_i$,
\item $\bm{a}_i =$ the position of $q_i$ from the center of mass of the magnet,
\item $a =$ the radius of magnet,
\item $\bm{\rho} =$ the position of the center of mass of the magnet from the
origin,
\item $\bm{a}'_i = \bm{a}_i + \bm{\rho}$,
\item $\bm{v}'_i = \bm{v}_i + \bm{V}$, where
\item $\bm{v}_i =$ the velocity of $q_i$ with respect to the center of mass of
the magnet.
\end{itemize}
If the (neutral and unpolarized) magnet is massive enough, $\bm{v}'_i$ can
be replaced with $\bm{v}_i$ in the Lagrangian. Also the current in the magnet is
assumed here to be positive, so a $q_i$ associated with its $\bm{v}_i$ is
positive, and negative charges do not appear in the Lagrangian when $\bm{V}$ is
absent.

When converting the sums to integrals, the following relationships are used.
\begin{itemize}
\item $q_i \rightarrow \lambda a d\phi$, where
\item $\lambda =$ the linear charge density of the positive current,
\item $\bm{a}_i \rightarrow a(cos\phi\bm{\hat{i}} + sin\phi\bm{\hat{j}})$,
\item $\bm{v}_i \rightarrow \bm{\omega}\times\bm{a} = \omega a\bm{\hat{\phi}}$, where,
\item $\bm{\hat{\phi}} = -sin\phi\bm{\hat{i}} + cos\phi\bm{\hat{j}}$.
\end{itemize}

\subsection{The general momentum of the point charge}\label{GM of PC}

This result will come out the same as that of Coleman and Van Vleck
\cite{Coleman}, their equation (22). I call the momentum of the point charge
$\bm{P}$. It is found by taking the gradient of the Darwin Lagrangian, $L$,
with respect to the velocity, $\bm{v}$, of the point charge. The equation for
the general momentum for the point charge is
\begin{eqnarray}
\label{dL/dv}
\bm{P} = \frac{\partial L}{\partial \bm{v}} &=&
m\bm{v}\left(1+\frac{v^2}{2c^2}\right)\\ \nonumber
&+& \frac{q}{8\pi\epsilon_o c^2}\sum_i\frac{q_i}{|\bm{r} + \bm{\rho} - \bm{a}_i|} \\ \nonumber
&\times&\left[\bm{v}_i + \frac{(\bm{v}_i\cdot(\bm{r} + \bm{\rho} -\bm{a}_i))}{|\bm{r} + \bm{\rho} -\bm{a}_i|^2}(\bm{r} + \bm{\rho} -\bm{a}_i)\right].
\end{eqnarray}
The first term on the right-hand side is the relativistic mechanical momentum
to order $c^{-2}$, and the second term is the electromagnetic momentum to the
same order of approximation.

In converting the summation above into an integral, it is convenient to minimize
the ``busyness'' of the math displayed by dropping $\bm{\rho$} from the
equations. This can be done because the length $\rho$ is negligible compared to
the length $r$ and can be negligible to $a$ if $M$, the mass of the
magnet, is sufficiently large. These
assumptions appear to be compatible with those of Coleman and Van Vleck.
Converting the sum in the above equation to an integral using the relationships
above, you obtain
\begin{eqnarray}
\label{canonical mo 1}
\bm{P} &=& m\bm{v}\left(1+\frac{v^2}{2c^2}\right) \\ \nonumber
&+& \frac{q\omega\lambda a^2}{8\pi\epsilon_o c^2}\left[\int_0^{2\pi}\frac{(-sin\phi\bm{\hat{i}}+cos\phi\bm{\hat{j})}d\phi}{(r^2 + a^2 + 2racos\phi)^{1/2}} \right. \\ \nonumber
&+& \left. \int_0^{2\pi}\frac{\bm{r}\cdot(-sin\phi\bm{\hat{i}}+cos\phi\bm{\hat{j}})(\bm{r} - a(cos\phi\bm{\hat{i}}+sin\phi\bm{\hat{j}}))d\phi}{(r^2 + a^2 + 2racos\phi)^{3/2}}\right].
\end{eqnarray}

There are four nonzero integrals over $\phi$ to evaluate. These integrals can
be approximated by noting that, consistent with the model of Shockley and James
and that of Coleman and Van Vleck, $r$ can be considered to be much larger
than $a$. This allows you to expand the denominators in powers of $a/r$ before
performing the integrations. With these approximations, the general momentum
of the point charge becomes
\begin{eqnarray}
\label{canonical mo 2}
\bm{P} &\approx& m\bm{v}\left(1+\frac{v^2}{2c^2}\right) \\ \nonumber
&+& \frac{q\omega\lambda a^2}{8\pi\epsilon_o c^2}\left[-\frac{\pi a}{r^2}\bm{\hat{j}} - \frac{3\pi a(\bm{\hat{r}}\cdot\bm{\hat{j}})}{r^2}\bm{\hat{r}} - \frac{\pi a(\bm{\hat{r}}\cdot\bm{\hat{j}})}{r^2}\bm{\hat{i}} + \frac{\pi a(\bm{\hat{r}}\cdot\bm{\hat{i}})}{r^2}\bm{\hat{j}}\right],
\end{eqnarray}
where $\bm{\hat{r}}$ is the unit vector in the $\bm{r}$ direction. Now $\bm{r}$
is a long position vector in the $-x$ direction before the disks are brought to
rest, so you might think you could replace $\bm{\hat{r}}$ to a good
approximation by $-\bm{\hat{i}}$. Also, replace $c^{-2}$ with $\mu_o\epsilon_o$.
When this is done, the above equation becomes
\begin{equation}
\label{canonical mo 3}
\bm{P} \approx  m\bm{v}\left(1+\frac{v^2}{2c^2}\right) - \frac{\mu_o q\omega\lambda a^3}{4 r^2}\bm{\hat{j}}.
\end{equation}
Next note the following.
\begin{eqnarray}
\label{magnetic moment}
(\omega\lambda a)(\pi a^2)\bm{\hat{j}} &=& IA\bm{\hat{j}} = IA(-\bm{\hat{i}}\times\bm{\hat{k}}) = \bm{\hat{r}}\times IA\bm{\hat{k}} \\ \nonumber
&=& \frac{\bm{r}}{r}\times\bm{\mathcal{M}},
\end{eqnarray}
where $\bm{\mathcal{M}}$ is the magnetic moment of the magnet and, as before,
$\bm{\hat{r}}$ has been identified with $-\bm{\hat{i}}$. This lets you express
Eq. (\ref{canonical mo 3}) as
\begin{equation}
\label{C&V eq 22}
\bm{P} \approx m\bm{v} - \frac{\mu_o q\bm{r}\times\bm{\mathcal{M}}}{4\pi r^3},
\end{equation}
which is Coleman and Van Vleck's equation (22) to the same degree of
approximation as they used. This result can also be written as.
\begin{equation}
\label{C&V eq 22 ala JJ}
\bm{P} = m\bm{v} + \mu_o\epsilon_o\bm{E}\times\bm{\mathcal{M}},
\end{equation}
where $\bm{E}$ is the electric field at the position of the magnetic moment.

\subsection{Motion of the center of energy and equation of motion of the magnet}

The center of energy, following Coleman and Van Vleck
\cite{Coleman}, is,
\begin{eqnarray}
\label{center of energy}
\frac{E\bm{X}}{c^2} &=& m\left(1 + \frac{v^2}{2c^2}\right)\bm{r} + M\left(1 + \frac{V^2}{2c^2}\right)\bm{\rho} \\ \nonumber
&+& \sum_i m_i\left(1 + \frac{v_i^2}{2c^2}\right)\bm{a}_i
\frac{q}{4\pi\epsilon_o c^2}\sum_i\frac{q_i}{|\bm{a}_i - \bm{r}|}\bm{r} \\ \nonumber
&+& \frac{1}{8\pi\epsilon_o c^2}\sum_{i\neq j}\frac{q_i q_j}{|\bm{a}_i - \bm{a}_j|}\bm{a}_i,
\end{eqnarray}
where $E$ here is the energy and $\bm{V}$ is the velocity of the magnet. (Note
that this equation has the units of mass times position rather than just
position due to the inclusion of electromagnetic energy.) Again following
Coleman and Van Vleck, I write the time derivative of the center of energy as
follows.
\begin{eqnarray}
\label{center of energy velocity}
&\textrm{}&\frac{1}{c^2}\frac{d(E\bm{X})}{dt} = m\left(1 + \frac{v^2}{2c^2}\right)\bm{v} + M\left(1 + \frac{V^2}{2c^2}\right)\bm{V} \\
&+& \sum_i m_i\left(1 + \frac{v_i^2}{2c^2}\right)\bm{v}_i \nonumber \\
&+& \frac{q}{8\pi\epsilon_oc^2}\sum_i\frac{q_i}{|\bm{r} - \bm{a}_i|}\left[\bm{v}_i
+ \frac{(\bm{v}_i\cdot(\bm{r} - \bm{a}_i))}{|\bm{r} - \bm{a}_i|^2}(\bm{r}
- \bm{a}_i)\right] \nonumber \\
&+&  \frac{1}{8\pi\epsilon_oc^2}\sum_{i\neq j}\frac{q_i q_j}{|\bm{a}_i - \bm{a}_j|}\left[\bm{v}_j
+ \frac{(\bm{v}_j\cdot(\bm{a}_i - \bm{a}_j))}{|\bm{a}_i - \bm{a}_j|^2}(\bm{a}_i - \bm{a}_j)\right], \nonumber
\end{eqnarray}
which has the units of momentum.

In order to evaluate Eq. (\ref{center of energy velocity}), the speeds of the
particles must be determined by the applied electric field. The
non-interacting masses $m'$, confined to a frictionless circular track,
experience an angle-dependent, radially-directed normal force, $N$, and an
electric force in the positive $x$ direction of magnitude $Eq'$ due to the
external charge, where $q'$ is the
charge on the mass $m'$. (Here, primed quantities are associated with the
magnet.) Let $a$ be the radius of the track, $\phi$ the azimuth angle that
locates the particle $m'$ at time $t$ with speed $v$, and $\phi_o$ the initial
azimuth angle where the particle has the initial speed $v_o$ at the time,
$t = 0$, when the electric field is applied. Taking $x = acos\phi_o$ as the
zero point of the electrical potential energy, the energy conservation equation
for a particle is
\begin{equation}
\label{energy}
m'\gamma c^2 = m'\gamma_o c^2 + q'Ea(cos\phi - cos\phi_o)
\end{equation}
where $\gamma$ and $\gamma_o$ are the Lorentz factors for $v$ and $v_o$. Solving
for $v$ by expanding its Lorentz factor, you get
\begin{equation}
\label{v squared}
v^2 = c^2 - \frac{c^2}{\gamma_o^2\left[1 + \frac{q'Ea}{m'\gamma_o c^2}(cos\phi - cos\phi_o)\right]^2}.
\end{equation}
Expanding $\gamma_o$ in terms of $v_o^2/c^2$ to first order, you find
\begin{equation}
\label{v}
v = v_o\left[1 + \frac{2q'Ea}{m'\gamma_o v_o^2}(cos\phi - cos\phi_o)\right]^\frac{1}{2},
\end{equation}
where $2q'Ea/(m'\gamma_o v_o^2)$ is assumed to be very small compared to 1. This
will be true for a weak electric field where particle accelerations are
small enough that radiation can be neglected.
Since $v = \omega a$ and $v_o = \omega_o a$ due to the circular motion, this
equation becomes
\begin{equation}
\label{omega}
\omega = \omega_o\left(1 + \frac{2q'E}{m'\gamma_o \omega_o^2 a}(cos\phi - cos\phi_o)\right)^\frac{1}{2},
\end{equation}
where the revolution is in the positive direction in the $x$-$y$ plane. (A
similar calculation was done by Boyer \cite{Boyer3}.)

The above is the angular motion of a single particle, beginning at $\phi_o$ with
speed $v_o$. To get the non-relativistic motion, all you have to do is set
$\gamma_o = 1$. Since the non-relativistic motion is sufficient to obtain the
result of Coleman and Van Vleck, this will be done in subsequent work in this
section.

To get the motion of a collection of particles moving around the track
and creating a uniform current, imagine inserting particles at $\phi_o = \pi/2$
one at a time in equal intervals with speed $v_o$ until the current is complete.
This process will add momentum to the system in the negative $x$ direction but
not in the $y$ direction. Also, ignore the Lorentz forces that arise to imitate
the disregard by Coleman and Van Vleck of how the system came to be. Eq.
(\ref{omega}) with $cos\phi_o = 0$ will now describe the motion of each
particle at a given angle $\phi$. The non-uniform angular motion will also
result in a non-uniform linear charge density, $\lambda$, the motion of which
constitutes the current. A short calculation shows, as expected,
$\lambda$ to vary inversely as $\omega$. That is,
\begin{equation}
\label{lambda}
\lambda = \lambda_o\left(1 + \frac{2q'E}{m'\omega_o^2 a}cos\phi\right)^{-\frac{1}{2}},
\end{equation}
where $\lambda_o$ is the linear charge density at $\phi = \pi/2$. This means
$ I = \lambda\omega a = \lambda_o\omega_o a$ and the evaluation of Eq.
(\ref{canonical mo 3}) is the same as before. (The current remains uniform as
required.) The general momentum is once again Eq. (\ref{C&V eq 22 ala JJ}).

The mass distribution in the magnet is also not uniform due to the non-uniform
angular speed. Like the linear charge distribution, mass becomes more
concentrated when the speed slows and less concentrated when the speed
increases. The linear mass density of the current particles is
\begin{equation}
\label{linear mass density}
d = d_o\left(1 + \frac{2q'E}{m'\omega_o^2 a}cos\phi\right)^{-\frac{1}{2}},
\end{equation}
where $d_o$ is the mass density at $\phi = \pi/2$.

The time derivative of the general momentum of Eq. (\ref{C&V eq 22 ala JJ})
is taken to be zero, as in the analysis of Coleman and Van Vleck (to their
degree of approximation), since the magnet is assumed to be neutral and
unpolarized and terms on the order of $v^2/c^2$ are neglected. (See their
equation (7)). This means, as Coleman and Van Vleck found in the equation below
their equation (22), the force on the charge is
\begin{equation}
\label{force on charge}
m\frac{d\bm{v}}{dt} = - \mu_o\epsilon_o\bm{E}\times\frac{d\bm{\mathcal{M}}}{dt}.
\end{equation}

Turning next to the time derivative of the velocity of the center of energy,
Eq. (\ref{center of energy velocity}),
The second sum has already been evaluated in the calculation of Eq.
(\ref{dL/dv}), resulting in Eq. (\ref{C&V eq 22 ala JJ}). The double summation
is zero due to the assumption on non-interaction, but the first sum, the kinetic
energy, does not vanish. The result is, to the same approximation as Coleman
and Van Vleck,
\begin{eqnarray}
\label{C of E velocity integral}
\frac{1}{c^2}\frac{d(E\bm{X})}{dt}
&=& m\bm{v} + \sum_im_i\frac{v_i^2}{2c^2}\bm{v}_i \\ \nonumber
&+& M\bm{V} - \frac{\mu_o q\bm{r}\times\bm{\mathcal{M}}}{4\pi r^3} \\ \nonumber
 &=& m\bm{v} + \sum_im_i\frac{v_i^2}{2c^2}\bm{v}_i + M\bm{V} + \mu_o\epsilon_o\bm{E}\times\bm{\mathcal{M}}.
\end{eqnarray}
Using Eq. (\ref{linear mass density}) and the relationships previously
listed to turn summations into integrals, the summation in the above equation
becomes
\begin{eqnarray}
\label{relativistic KE}
\sum_im_i\frac{v_i^2}{2c^2}\bm{v}_i
&\rightarrow& \frac{a^4d_o\omega_o^3}{2c^2} \int_0^{2\pi}\\ \nonumber
&\times& \left(1 + \frac{2q'E}{m'\omega_o^2 a}cos\phi\right)\bm{\hat{\phi}}d\phi \\ \nonumber
&=& \frac{a^3\lambda_o E\omega_o}{c^2}\bm{\hat{j}}\int_0^{2\pi}cos^2\phi d\phi = \frac{\mu_oq\omega_o\lambda_oa^3}{4r^2} \\ \nonumber
&=& - \mu_o\epsilon_o\bm{E}\times\bm{\mathcal{M}},
\end{eqnarray}
Then, using Eqs. (\ref{force on charge}) and (\ref{relativistic KE}), the
derivative of the velocity of the center of energy becomes
\begin{eqnarray}
\label{C of E = 0, slot model}
\frac{1}{c^2}\frac{d^2(E\bm{X})}{dt^2} &=& 0 = - 2\mu_o\epsilon_o\bm{E}\times\frac{d\bm{\mathcal{M}}}{dt} \\ \nonumber
&+& M\frac{d\bm{V}}{dt} + \mu_o\epsilon_o\bm{E}\times\frac{d\bm{\mathcal{M}}}{dt} \\ \nonumber
&\rightarrow& M\frac{d\bm{V}}{dt} = \mu_o\epsilon_o\bm{E}\times\frac{d\bm{\mathcal{M}}}{dt},
\end{eqnarray}
which is the same result obtained by Coleman and Van Vleck. (For those possibly
confused by the sign difference between the above equation and their equation
(27), note that $\bm{E} = -1/(4\pi\epsilon_o)q/r^2\bm{\hat{i}}$ since $\bm{r}
= -r\bm{\hat{i}}$.)

However, there is a serious oversight in the derivation of the above equation.
For one thing notice how you can create a uniform current corresponding to Eq.
(\ref{omega}) by adding particles one at a time at equal intervals at an
\textit{arbitrary} angle $\phi_o$. Instead
of an initial momentum of zero in the $y$ direction and $-Nm'v_o$ in the $x$
direction, there would be a non-zero momentum of $Nm'v_ocos\phi_o$ in the $y$
direction and $-Nm'v_osin\phi_o$ in the $x$ direction added to the system, where
$N$ is the number of inserted particles. The consequences of this observation
is discussed in the next section.

\section{A critical examination of the Coleman-Van Vleck result}

The Coleman-Van Vleck result \cite{Coleman} would seem to confirm the suspicion
of Shockley and James \cite{Shock} that hidden momentum can result from
relativistic kinetic energy effects. Thus it would seem that the
force applied to the external charge, Eq. (\ref{force on charge}), is balanced
by the opposite force applied to the magnet, Eq. (\ref{C of E = 0, slot model}),
and momentum is conserved.

The claim by Coleman and Van Vleck that the magnet in their model experiences
an equal and opposite force to that of the point charge is expressed in their
equation (15). There is, in my view, an error in the interpretation of this
equation. This equation is, in their notation,
\begin{equation}
\label{CVV 15}
m\frac{d^2\bm{r}}{dt^2} = -M\frac{d^2\bm{X}_m}{dt^2},
\end{equation}
where $m$ and $\bm{r}$ are the mass and position of the point charge, $M$ is
the mass of the magnet, and $\bm{X}_m$ is the magnet's center of energy. The
terms that go into $M$ and $\bm{X}_m$ are terms that could apply to the
particles in the
magnet and/or the electromagnetic field. Therefore, I argue that the term on
the right hand side of the above equation could just as well correspond to
the loss of electromagnetic momentum due to the decay of the field as to a
force on the magnet. The original application of an electric field to the magnet
resulted in the magnet obtaining equal and opposite amounts of electromagnetic
and mechanical momentum. When the magnetic field decays, as in the SJ model, the
momentum in the field is converted to mechanical momentum of the charge, and the
system, which was moving in its original rest frame, now has zero mechanical
momentum.

Coleman and Van Vleck calculate the force presumably applied to the magnet in
their equation (26). This result comes from the time rate of change of the
center of energy, their equation (11). I rewrite this equation below in their
notation, separating the terms involving the point charge.
\begin{eqnarray}
\label{CVV 11}
\frac{d}{dt}(E\bm{X}) &=& m(1 + \frac{v^2}{2c^2})\bm{v} + \sum_a m_a
(1 + \frac{v_a^2}{2c^2})\bm{v}_a \\ \nonumber &+& \frac{e}{c^2}\sum_a\frac{e_a}{|\bm{r} - \bm{r}_a|}
\left[\bm{v} + \frac{(\bm{r}-\bm{r}_a)(\bm{v}\cdot(\bm{r}-\bm{r}_a))}{|\bm{r}-\bm{r}_a|^2}\right] \\ \nonumber
&+& \frac{e}{c^2}\sum_a \frac{e_a}{|\bm{r}-\bm{r}_a|}
\left[\bm{v}_a + \frac{(\bm{r}-\bm{r}_a)(\bm{v}_a\cdot(\bm{r}-\bm{r}_a))}{|\bm{r}-\bm{r}_a|^2}\right] \\ \nonumber
&+& \frac{1}{2c^2}\sum_{a\neq b} \frac{e_a e_b}{|\bm{r}_a-\bm{r}_b|}
\left[\bm{v}_b + \frac{(\bm{r_a}-\bm{r}_b)(\bm{v}_b\cdot(\bm{r}_a-\bm{r}_b))}{|\bm{r}_a-\bm{r}_b|^2}\right] \\ \nonumber
&\approx& m\bm{v} + M\frac{d\bm{X}_m}{dt}.
\end{eqnarray}
Here, $m$, $\bm{v}$, $\bm{r}$, and $e$ refer to the mass, velocity, position,
and charge, respectively, of the point charge. The subscripted quantities refer
to the particles comprising the magnet. The last two terms on the right hand
side of the above equation are gotten from their equation (14). Their equation
(25) is
\begin{equation}
\label{CVV 25}
M\frac{d\bm{X}_m}{dt} = \bm{P}_m - \frac{e}{c^2}\sum_a e_a\frac{(\bm{r}\cdot\bm{v}_a)\bm{r}_a}{r^3},
\end{equation}
where $\bm{P}_m$ is the momentum of the isolated magnet and the second term on
the right is what remains of the magnetic portion of Eq. (\ref{CVV 11}) after
approximations have been made and the velocity of the external charge has been
taken as
zero ($\bm{v} = 0$). This means they have applied the center of energy theorem
without taking into account the original momentum involved in the assembly of
the system. At the beginning there were components destined
to be parts of this system which had zero linear momentum. Assuming for the
present case that no external agent kept the system stationary when it was put
together, it would contain the mechanical momentum due to the Lorentz forces
that arose during assembly. It also has the equal and opposite electromagnetic
momentum in the fields.

If the system of Coleman and Van Vleck has hidden momentum equal and opposite
to the electromagnetic momentum, when the magnetism dies out (for example, by
the brake mechanism of Shockley and James) the electromagnetic momentum is
converted to the mechanical momentum of the charge and the hidden momentum to
the mechanical momentum of the magnet. They are equal and opposite, but the
system is still moving with respect to its original rest frame due to the
mechanical momentum imparted by the Lorentz forces. There is nothing to balance
this momentum such that momentum is not conserved: This reference frame
originally had zero momentum and now it has a net momentum. The paradox of
Shockley and James is therefore not a paradox at all; hidden momentum, if it
exists, is.

A side issue is the presence of electromagnetic angular momentum in the
single-charge Coleman-Van Vleck model. A charge-magnet system has both linear
and angular momentum in its fields \cite{Furry}. For the model of Shockley and
James, the presence of two equal and opposite charges at equal distances from
the magnet results in zero angular field momentum, such that when the magnetic
field decays, the equal impulses applied to the charges result in zero
mechanical angular momentum. In the Coleman-Van Vleck case, however, there is a
net angular momentum in the fields. When the magnetic field decays this results
in mechanical angular momentum due to the impulse received by the single charge.

The circular frictionless track model based on the work of Coleman and Van Vleck
and their footnote 9, appears in a more accessible version in Babson
\textit{et al} \cite{Babson} in their ``cleanest example of hidden momentum''.
There is an error common to both these models. Other than ignoring the Lorentz
forces that arise during the assembly of the systems, also ignored is the
momentum transfer between the particles and the track they follow. The result
is the system starts out already containing  momentum in the frame of reference
in which it is viewed. As pointed out in the above section,
you can start out with the magnet
consisting of charged particles traveling freely on a circular track having any
momentum whatsoever and still get the results of Coleman and Van Vleck. This is
because you are either viewing the magnet in its new rest frame, or the magnet
is held stationary and momentum is added to its environment. The hidden momentum
calculated in the above section is only seen in the frame of reference at rest
with the track.

You might think that this artificial situation can be resolved by starting out
with a magnet consisting of non-interacting charge carriers that form a uniform
current before the electric field is applied. However, as will be shown below,
the Coulomb forces cannot transfer net mechanical momentum to either the
circular track model or the Babson \textit{et al} model in a direction
perpendicular to the field. In all the analyses that follow, it will be assumed
that an external agent provides forces to counter the (magnetic) Lorentz forces
so that the effects of the Coulomb forces can be isolated. This mimics the
disregard of the Lorentz forces in the assembly of the systems.

It is perhaps clearest to see this oversight by looking at the example of Babson
\textit{et al} (``cleanest example of hidden momentum''). In one scenario,
their model is a rectangular frictionless track containing non-interacting
positive particles that nevertheless respond to a uniform electric field
applied across the track. Imagine the bottom of the track defining the $x$ axis
and the left side the $y$ axis with the origin of the axes at the lower left
corner of the track. The corners of the rectangular track are curved so that
the particles can move around the track without making a normal collision with
a side (see Fig. 4).

The electric field is in the positive $y$ direction such that particles
accelerate in that direction (``up'') on the left side of the track, turn the
corner and move with constant speed in the positive $x$ direction (to the right)
at the top of the track, decelerate while moving in the negative $y$ direction
(``down'') on the right side of the track, then finally move to the left along
the bottom of the track at a speed less than what they had at the top of the
track. The number of particles on the bottom of the track in their model is
taken to be larger than that on the top by an amount necessary to maintain a
uniform current.

They examine two versions of their particle scenario: a Newtonian
version where the mass of the particles does not depend on their speed and a
relativistic version that includes the relativistic increase of mass with speed.
In the latter version the total momentum of the particles at
the top of the track to the right is purported to be greater than that of the
larger number of particles at the bottom of the track to the left due to this
relativistic effect that increases the mass to first order by a factor of
$1 + v^2/2c^2$. This
is the source of the hidden momentum the track is supposed to hold. There is
supposedly a net (hidden) momentum in the positive $x$ direction equal and
opposite to the electromagnetic field momentum that results from the presence of
the electric and magnetic fields, presumably conserving momentum.

\begin{figure}[ht!]
\label{Babson model}
\centering
\includegraphics[width=3in]{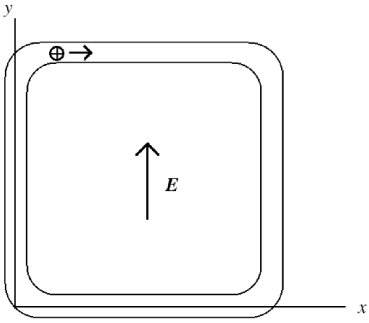}
\caption{The Babson \textit{et al} model.}
\end{figure}

The error arises when Babson \textit{et al} do not consider how the system
evolves under the action of Coulomb forces after the electric field is applied.
For simplicity and clarity, imagine their track is fixed so it can't move
parallel to the $y$ direction but can slide frictionlessly parallel to the
$x$ direction. To illustrate the problem involving the momentum exchange at the
corners of the track, follow a single particle after the electric field has been
applied. Have it start at the origin moving with a small initial speed in the
positive $y$ direction (up the left side of the track). It accelerates until it
reaches the upper left corner at which point it is deflected to the right. The
total momentum in the $x$ direction was zero to begin with, and now it is still
zero since the track must recoil to the left.
(There is, of course, momentum transfer to the track in the positive $y$
direction. If the track is held stationary in that direction, this momentum
will be absorbed by the environment.)

Momentum transfer between the particle and track also occurs at the other three
corners while the total momentum in the $x$ direction remains zero. Finally the
initial situation is regained with the particle moving up the left side of
the track. Note that it doesn't matter whether or not the mass of the particle
is increased by relativistic effects at the top of the track and that the
electric field does not change the momentum of the system in a direction
perpendicular to itself (parallel to the $x$ direction). Also note that, just
as in the circular track model, you
can start the particle anywhere and begin with a system containing arbitrary
momentum. However, at any given time the track and the particle can each have a
non-zero momentum, but these will be equal and opposite to maintain momentum
conservation.

In the scenario involving the frictionless track version of the Darwin
Lagrangian calculation, the particles were introduced at $\phi = \pi/2$ with an
initial speed in the negative $x$ direction at equal time intervals to create a
uniform current. No net momentum in the $y$ direction is introduced to the
magnet by this procedure. The Lagrangian calculation for that situation
resulted in a net momentum in a direction perpendicular to the electric field
(parallel to $y$). Why is that? To find out you can once again look at the
analogous Babson \textit{et al} model.

Issue four particles from the lower left corner at equal time intervals with +1
unit of momentum in the $y$ direction and have the track held immobile so that
any momentum transferred to the track goes into its environment. Let the time
interval be such that at any given time each particle is on a different side
of the track. Due to the action of a constant Coulomb force in the positive $y$
direction, say the particles gain +1 unit of momentum in the $y$
direction on the left (speeding up) and right (slowing down) sides of the track.
The particles will collide elastically with the corners at different times but
will always be separated from each other in time by the chosen time interval.

After $N$ (where $N \geq 3$) collisions have taken place at the top left corner
of the track, $N -1$ have taken place at the upper right corner, $N-2$ at the
lower right corner, and $N-3$ at the lower left corner. Looking only at momentum
parallel to the $x$ direction, a collision at
the upper left will impart a momentum of -2 units to the track, one at the
upper right will impart +2 units, one at the lower right +1 unit, and one at
the lower left -1 unit. So the net momentum imparted to the track (and
subsequently transferred to its environment) after $N$ collisions at the upper
left is $-2N + 2(N-1) + (N - 2) - (N - 3) = -1$ unit in the $x$ direction.

The track's environment gained -1 unit of momentum in the $x$ direction after
the first two sets of collisions, and its momentum does not change afterwards.
However, there is always a particle at the top of the track with a momentum of
+2 and one at the bottom with a momentum of -1, so the total momentum of the
system, particles plus track plus environment, is always zero. Adding more
particles in a way to keep the current uniform will not change this picture
qualitatively: The environment will contain momentum in the negative $x$
direction while the particles will contain and equal and opposite momentum in
the positive $x$ direction.

The momentum of the system is zero, but if you view it from the track's rest
frame, it appears that there is a net (hidden) particle momentum of $n/4$ units
in the positive $x$ direction, where $n$ is the number of particles. No
relativistic effects are needed. The calculation performed earlier for the
Coleman and Van Vleck model with the frictionless track does just that. The
calculation is in effect performed in the rest frame of the track.

The above scenario is, of course, quite contrived. A more realistic scenario is
one where the particles are initially moving around the track with the same
speed equally spaced. The total momentum of the particles is zero. Once the
electric field is applied, particles on the left side of the track moving up
will gain momentum and particles on the right moving down will lose momentum. As
a result of the particles having different speeds, the faster ones will pass the
slower ones, assuming the particles are non-interacting as was done by both
Coleman and Van Vleck and by Babson \textit{et al}.

So, although you start out with a uniform current, this will change due to the
effect of the electric field. As the system evolves, the total momentum of the
particles moving right
at the top of the track will be equal and opposite to that of the particles
moving to the left at the bottom due to equal and opposite collisions at the
corners. No hidden momentum results.

This situation is also true for a circular track if you start out
with a uniform current instead of artificially creating one by introducing
particles one by one. From Eq. (\ref{omega}) you find, approximately,
\begin{equation}
\label{dt}
dt = \frac{d\phi}{\omega} = \frac{1}{\omega_o}\left[1 - \frac{q'E}{m'\omega_o^2 a}(cos\phi - cos\phi_o)\right]d\phi.
\end{equation}
When you integrate this from $\phi_o$ to $\phi_o + 2\pi$, you find that the
period of a particle starting out at $\phi_o$ with angular speed $\omega_o$ when
the electric field is applied is about
\begin{equation}
\label{T}
T = T_o(1 + \frac{q'E}{m'\omega_o^2 a}cos\phi_o),
\end{equation}
where $T_o = 2\pi/\omega_o$ is the original period. In general each particle
will have a different average angular speed, $\bar{\omega} = 2\pi/T$. In fact,
after a time
\begin{equation}
\label{t}
t = \frac{2\pi - (\phi_2 -\phi_1)}{\omega_2 - \omega_1} = \frac{m'\omega_o a}{q'E}\left[\frac{2\pi - (\phi_2 - \phi_1)}{cos\phi_1 - cos\phi_2}\right]
\end{equation}
goes by, a faster particle originally at $\phi_2$ will catch up to a slower one
originally at $\phi_1$.

A way to get around this unphysical result is to assume the particles are
interacting such that the speed of the particles remains the
same (or, equivalently, the non-relativistic linear mass density remains
uniform). Due to the uniform speed, the relativistic mass effect will not work
to produce hidden momentum. However there will still be relativistic momentum
due to relativistic effects involving pressure. Babson \textit{et al} address
this in their scenario of an incompressible and frictionless fluid flowing
through a tube shaped like their track.

Babson \textit{et al} use the relativistic expression for the momentum density
in an incompressible frictionless electric fluid due to pressure
($= \gamma^2 P\bm{v}/c^2$ where $P$ is the pressure and $\bm{v}$ is the fluid
velocity) to show that there is a momentum difference, $(\gamma/c)^2(P_T -
P_B)vlA$, between the top and bottom of the tube due to the pressure
difference, $P_T - P_B$, produced by the electric field, where $v$
is the fluid speed, $l$ is the length of the track segments, and $A$ is the
cross-sectional area. This is certainly the case, but to say this is the
net (hidden) momentum in the system runs into difficulties.

There is a simple thought experiment that is apropos to this situation. Instead
of having a uniform electric field exert a force on a charged incompressible
circulating fluid, say you apply a uniform gravitational field to an
uncharged incompressible fluid. There is, of
course, no electromagnetic momentum in the system, yet the fluid will still
acquire the relativistic momentum that appears in the Babson \textit{et al}
example due to the pressure created by the gravitational force. Although it is
unclear to me how a gravitational field might contain hidden momentum
(apparently, there has to be a Maxwell's theory of gravity of some sort), such
a hidden momentum has been proposed \cite{Hnizdo}. Covering this is beyond the
scope of this paper, but I argue below there is no need for hidden
momentum in the gravitational field.

I look upon this situation as a case of momentum flow due to energy flow
\cite{Planck}.
As the fluid flows in the
direction of the gravitational field, it gains energy due to the increasing
pressure. This energy then flows horizontally after it encounters the
corner toward which it is moving. After encountering the next corner it flows
counter to the gravitational field, losing energy. In the Newtonian view
the energy lost or gained by the fluid is due to the gain or loss, respectively,
of energy stored in the gravitational field.

Looking at the momentum, there is the same situation here regarding interaction
between track and particles as in the non-interacting particle scenarios. The
momentum transfer between the fluid and tube at the top left corner is equal
and opposite to that at the top right corner (similarly for the bottom corners).
The corner collisions produce a stress in the tube -- a tension
equal and opposite to the pressure in the fluid. Due to time differences
between when a particular slug of fluid collides with the upper left corner
and the upper right corner, there is a flow of stress in the tube, resulting
in a steady state flow after the electric field reaches full strength. This
stress counters the pressure in the fluid so that there is no net momentum
flow. Planck \cite{Planck} conjectured that in relativity theory
when there is energy flow there is also momentum. If this is true, there
can be no net momentum in the system since the net energy flow is zero. Momentum
flow -- that is momentum traveling through non-moving media -- is important in
other paradoxes, for example, the paradox of Trouton and Noble
\cite{Trouton,Laue,redfern3}.

Summing up, the problem with both the model of Babson \textit{et al} and the
application of the circular frictionless track to the model of Coleman and Van
Vleck is the same. Both ignore the momentum interactions within their models,
and both ignore the Lorentz forces that arise when the models were
formed or assembled. As has been shown, the Coulomb forces due to the
application of an electric field cannot add net mechanical momentum to the
models (plus environment if appropriate) in a direction perpendicular to the
electric field.

When the electric field is turned off in the case of the incompressible fluid,
the pressure in the fluid and stress in the tube will disappear as will the
relativistic momentum. Both the internal energy flow and the consequent
momentum flow will also disappear, and there would be no contribution to the
motion of the track from this change, as the thought experiment with the
gravitational field shows. (Of course, in all models you will get motion from
Lorentz forces when the electric field is removed.) You might indeed use
Boyer's term ``internal momentum'' for this relativistic momentum \cite{Boyer5},
but it has no effect on the motion of the magnet as the \textit{net momentum
flow in the direction perpendicular to the electric field is zero before,
during, and after the field application}. Hence this is different from the
concept of hidden momentum: The net hidden momentum in the particle scenarios
(if it existed) can affect the motion of the track, but this form of momentum
cannot.

A complicating factor is an erroneous assumption made by Coleman and Van Vleck
(and implicitly by Babson \textit{et al}) with regard to their non-interacting
particle scenarios. They assumed that the polarization
of the magnet is negligible due to the distance
between the magnet and the external charge when they implemented the Darwin
Lagrangian, but this fails to appreciate the polarization of the charge in the
magnet due to the non-uniform charge pattern (for example, more charge on the
bottom of the Babson \textit{et al} track than on the top).

If you examine the special case of a single particle orbiting a circular track
and calculate the effective electric dipole by integrating
$\bm{a}dq' = \bm{a}(a\lambda d\phi)$ over the loop while making the
assumption that $2q'E/m'\omega_o^2 a$ in Eq. (\ref{omega}) is small compared to
1, you get $-(q'E)/(m'\omega_o^2 a)(aq'/2)\bm{\hat{i}}$. The magnitude of the
effective dipole is equal to the ratio of the electric force on the charge
element to the mechanical centripetal force on that element times a dipole
moment of $aq'/2$. The effective dipole field in concert with the magnetic
dipole field will contribute a small amount to the electromagnetic field
momentum -- about one-sixth of that due to the interaction of the magnetic
dipole field with the electric field of the external charge if you approximate
the dipoles with spheres of radius $a$ containing uniform fields inside and
dipolar fields outside. Of course, the incompressible fluid scenario does not
exhibit this defect.

\section{A magnetic dipole in a uniform electric field}

In this section I will show that there is no hidden momentum when a magnet is
formed in a uniform electric field, using as the model a spherical shell
containing the uniform field with a magnet at its center. First I will calculate
the mechanical momentum imparted to the shell as the magnetic dipole is formed
and then show that the electromagnetic field momentum is equal and opposite to
this. No hidden momentum is necessary to conserve linear momentum.

\subsection{Mechanical momentum in the sphere-magnetic dipole system}

A uniform Coulomb electric field directed in the positive $y$ direction, given
by
\begin{equation}
\label{E_o}
\bm{E}_o = \frac{1}{4\pi\epsilon_o}\frac{p}{a^3}\bm{\hat{j}},
\end{equation}
will exist inside a spherical shell of radius
$a$ with an angular surface charge distribution given by \cite{Jackson}
\begin{equation}
\label{surface charge 2}
\sigma = -\frac{3p sin\theta sin\phi}{4\pi a^3},
\end{equation}
where $p$ is the dipole moment of the shell.
In this equation $\theta$ and $\phi$ are the polar and azimuth angles,
respectively, of a system of spherical coordinates with the origin at the center
of the sphere. Outside the sphere there is a dipole electric field with dipole
moment $p$.

Imagine that such a shell has a magnetic dipole formed at its center
pointing in the positive $z$ ($\bm{\hat{k}}$) direction, increasing uniformly
from zero to $\mathcal{M}\bm{\hat{k}}$. This could be done, for example, by
having two small nested spherical shells at the origin containing equal and
opposite surface charge (varying as $sin\theta$) with virtually no space
between them. Torques provided by an external agent could set the spheres
rotating about the $z$ axis, gradually picking up speed. The torques could be
equal and opposite, as would those arising from Lenz' law, so that no net
angular momentum is imparted.

The vector electromagnetic potential of the magnetic dipole at the location of
the shell is \cite{Reitz}
\begin{equation}
\label{vector potential}
\bm{A} = \frac{\mu_o}{4\pi}\frac{\mathcal{M}\bm{\hat{k}}\times\bm{\hat{r}}}{a^2} =
\frac{\mu_o}{4\pi}\frac{\mathcal{M}}{a^2}sin\theta(-sin\phi\bm{\hat{i}} +
cos\phi\bm{\hat{j}}).
\end{equation}
The increasing magnetic flux in the shell will induce a Faraday electric field
given by \cite{Reitz}
\begin{equation}
\label{Faraday E}
\bm{E} = - \frac{\partial\bm{A}}{\partial t} =
- \frac{\mu_o}{4\pi}\frac{\dot{\mathcal{M}}}{a^2}sin\theta(-sin\phi\bm{\hat{i}} +
cos\phi\bm{\hat{j}}),
\end{equation}
where $\dot{\mathcal{M}}$ is the time derivative of the magnetic moment.

This electric field will exert a force on each charge element on the
shell. The charge on an element of area $dA$ is given by
\begin{eqnarray}
\label{charge element}
dq &=& \sigma dA = -\frac{3p sin\theta sin\phi}{4\pi a^3}a^2 sin\theta d\theta d\phi \nonumber \\
&=& -\frac{3p sin^2\theta sin\phi}{4\pi a}d\theta d\phi.
\end{eqnarray}
An element of force is then
\begin{equation}
\label{force element}
d\bm{F} = \bm{E}dq = \frac{3\mu_o p\dot{\mathcal{M}}}{16\pi^2 a^3}sin^3\theta(-sin^2\phi\bm{\hat{i}}
+ sin\phi cos\phi\bm{\hat{j}})d\theta d\phi.
\end{equation}
Integrating the above equation over $\phi$ from 0 to $2\pi$ gets rid of the
$\bm{\hat{j}}$ term, whereas the integral of $sin^2\phi$ results in $\pi$. The
integral over $\theta$ of $sin^3\theta$ from 0 to $\pi$ is 4/3. The force
exerted on the shell due to the formation of the magnetic dipole is therefore
\begin{equation}
\label{Faraday force}
\bm{F} = -\frac{\mu_o p \dot{\mathcal{M}}}{4\pi a^3}\bm{\hat{i}}.
\end{equation}
(There is no Lorentz force on the magnetic dipole.) Since the time rate of
change of linear momentum equals the force, the mechanical momentum acquired by
the shell is
\begin{equation}
\label{mech momentum}
\bm{P}_{mech} = -\frac{\mu_o p \mathcal{M}}{4\pi a^3}\bm{\hat{i}}
= -\epsilon_o\mu_o\bm{E}_o\times\bm{\mathcal{M}},
\end{equation}
using Eq. (\ref{E_o}) to get the right-hand result.

\subsection{The electromagnetic momentum of the shell-magnetic dipole system}

Although point magnetic dipoles don't exist in nature, the magnetic field of
such an object is given by \cite{Jackson4}
\begin{equation}
\label{dipole B}
\bm{B} = \frac{\mu_o}{4\pi r^3}[3(\bm{\mathcal{M}}\cdot\bm{\hat{r}})\bm{\hat{r}} - \bm{\mathcal{M}}]
+ \frac{2}{3}\mu_o\bm{\mathcal{M}}\delta^3(\bm{r}),
\end{equation}
where $\delta^3(\bm{r})$ is the Dirac delta function. This magnetic field should
be applicable for a very small magnet inside the spherical shell. The creation
of this dipole imparts a momentum to the shell given by Eq.
(\ref{mech momentum}). The
magnetic dipole rests at the center of the sphere in a uniform electric field
given by Eq. (\ref{E_o}). The electromagnetic momentum is calculated by
Eq. (\ref{em momentum}) which inside the sphere is given by
\begin{flalign}
\label{EM momentum integral}
\bm{P}_{em} &= \frac{\mu_o \mathcal{M} E_o}{4\pi}\int_0^a \int_0^\pi \int_0^{2\pi}
\left [\frac{3 cos^2\theta - 1}{r^3}\bm{\hat{i}} \right. &&\nonumber \\
&- \left. \frac{3 sin\theta cos\theta cos\phi}{r^3}\right]
r^2 sin\theta dr d\theta d\phi + \frac{2}{3}\epsilon_o\mu_o \mathcal{M} E_o \bm{\hat{i}}.&&
\end{flalign}
The  integrals over the angles in Eq. (\ref{EM momentum integral}) are zero,
leaving the only contribution to the momentum from the Dirac delta term. The
electromagnetic momentum inside the shell is therefore
\begin{equation}
\label{Pem inside}
\bm{P}_{em-in} = \frac{2}{3}\epsilon_o \mu_o \bm{E}_o\times\bm{\mathcal{M}}.
\end{equation}
Outside the sphere the electric field is that of a dipole, given by
\begin{equation}
\label{E outside}
\bm{E} = \frac{1}{4\pi\epsilon_o r^3}[3(\bm{p}\cdot\bm{\hat{r}})\bm{\hat{r}} - \bm{p}].
\end{equation}
The electromagnetic momentum outside the shell is
\begin{flalign}
\label{Pem outside integral}
\bm{P}_{em-out} &= -\frac{\mu_o \mathcal{M} p}{16\pi^2}\int_V\frac{1}{r^6}\left(3 cos\theta \bm{\hat{r}}\times\bm{\hat{j}} \right. && \nonumber \\
&+ \left. 3 sin\theta sin\phi\bm{\hat{k}}\times\bm{\hat{r}} + \bm{\hat{i}}\right)dV && \nonumber \\
&= -\frac{\mu_o \mathcal{M} p}{16\pi^2}\int_0^a \int_0^\pi \int_0^{2\pi}\frac{1}{r^6}\left(-3 cos^2\theta\bm{\hat{i}}\right. && \nonumber \\
&+ \left. 3 cos\theta sin\theta cos\phi\bm{\hat{k}} - 3 sin^2\theta sin^2\phi\bm{\hat{i}} \right. && \nonumber \\
&+ \left. 3 sin^2\theta sin\phi cos\phi\bm{\hat{j}} + \bm{\hat{i}}\right)r^2 sin\theta dr d\theta d\phi. &&
\end{flalign}
The integrals over $\phi$ involving $sin\phi cos\phi$ from 0 to $2\pi$ are zero
and that over $sin^2\phi$ is $\pi$. The other integrals over $\phi$ are $2\pi$.
Hence the electromagnetic momentum outside the shell becomes
\begin{flalign}
\label{Pem outside reduced integral}
\bm{P}_{em-out} &= - \frac{\mu_o \mathcal{M} p\bm{\hat{i}}}{16\pi}\int_a^\infty
\frac{dr}{r^4}\int_o^\pi \left(-6 cos^2\theta sin\theta + 2 sin\theta \right. &&\nonumber \\
&- \left. 3 sin^3\theta\right)d\theta = \frac{1}{3}\frac{\mu_o \mathcal{M} p}{4\pi a^3}\bm{\hat{i}}.&&
\end{flalign}
Therefore
\begin{equation}
\label{Pem outside result}
\bm{P}_{em-out} = \frac{1}{3}\epsilon_o\mu_o\bm{E}_o\times\bm{\mathcal{M}}.
\end{equation}
Clearly the sum of Eqs. (\ref{mech momentum}), (\ref{Pem inside}), and
(\ref{Pem outside result}) is zero. There is no hidden momentum necessary to
achieve momentum conservation.

\section{A review of certain other work on hidden momentum}

One of the referees suggested that I include reviews of rather recent work --
some of which predated the submission of this manuscript. This was a good
suggestion, and it led me to other physics literature that was referenced
in these papers that I felt also needed to be commented on. The following is the
result of that effort.

Boyer \cite{Boyer6} argued that an Amperian current loop (as a model of a
neutron) experiences a force while passing a charged wire; that is, traveling
through an electrostatic field. In response Aharonov \textit{et al} \cite{APV}
claimed that Boyer overlooked the role of hidden momentum. This question is
central to the Aharonov-Casher effect \cite{AC}, where there is a phase
difference between a neutron passing on one side of the wire compared to one
passing on the other side. Aharonov \textit{et al} argue that the time rate of
change of the hidden momentum of the neutron counteracts the force
[$(\bm{p}\cdot\nabla)\bm{E}$ where $\bm{p}$ is the induced electric dipole
moment and $\bm{E}$ is the electric field] acting on the neutron such that the
total force on the neutron is zero. Here I intend to show there is no hidden
momentum in the neutron and it does experience a force in general while moving
in an electric field.

First, Aharonov \textit{et al} make the same error as others who examine
charge-magnetic dipole interactions. There \textit{is} momentum, electromagnetic
momentum,  in such a system when it is at rest. As argued here, the
application of an electric field to a magnet results in both mechanical and
electromagnetic momentum. If the neutron can be modeled as a tiny current loop,
you can envision how it comes to rest in an electric field as follows.

The neutron enters the region of the field with mechanical momentum. As it
progresses through the field, Lorentz forces impart additional mechanical
momentum as an equal and opposite amount of momentum is stored in the
electromagnetic field. If the additional mechanical momentum is just the right
amount to bring the neutron to rest in the lab frame, there will be an amount
of electromagnetic momentum associated with the neutron, but the mechanical
momentum will be zero; the missing momentum has been transferred to the
environment (for example, to the wire) by field interaction.

Whether or not the
neutron is brought to rest, the system involved consists of the source of the
electric field and the neutron -- not just the neutron alone. This is what
Aharonov \textit{et al} fail to take into account in their argument that a
neutron at rest in an electric field must contain hidden mechanical momentum.
With no hidden momentum in a neutron modeled as a current loop, you can
calculate the force acting on the neutron as Boyer did.

Vaidman \cite{Vaidman} discussed the torque and force on a magnetic dipole.
The situation addressed was that of a current loop with its magnetic moment in
the positive $x$ direction lying in the $y$-$z$ plane. There is a uniform
electric field in the positive $z$ direction. The loop is at rest in the $S'$
frame and moving in the positive $x$ direction in the $S$ frame. In the $S$
frame there is no magnetic field and no torque observed on the loop. However,
in the $S'$ frame traveling with the loop there is a magnetic field, Lorentz
forces acting on the current in the loop, and a subsequent torque given by
\begin{equation}
\label{torque on loop}
\bm{\tau} = \bm{\mathcal{M}}\times\bm{B}' = -\mathcal{M}B'\bm{\hat{k}},
\end{equation}
where $\bm{B}' = \gamma(v/c^2)E\bm{\hat{j}} = B_{y'}\bm{\hat{j}}$ is the
magnetic field in the $S'$
frame. This poses a paradox much like that of Mansuripur \cite{Mansuripur}.
Vaidman considers the work of Namias \cite{Namias} and Bedford and Krumm
\cite{Bedford} as pertaining to the resolution of the paradox. The model used
by Namias for the loop involves charges shielded by conducting material. The
model of Bedford and Krumm is like that of Coleman and Van Vleck \cite{Coleman},
where the current charges are non-interacting but exposed to the electric field.
Vaidman then proposes his own resolution based on a model like that of Babson
\textit{et al}, where there is a charged, incompressible fluid in a circular
tube. Here I will address the models where the charges are exposed to the
electric field, since they have hidden momentum-like mechanisms.

My resolution is much like that I presented concerning the paradox of Mansuripur
\cite{redfern1}. When the dipole is at rest in the $S$ frame the external
electric field and magnetic field of the loop contain linear electromagnetic
momentum given by \cite{Furry}
\begin{equation}
\label{p of loop}
\bm{P} = \bm{E}\times\bm{\mathcal{M}}/c^2 = [(E\mathcal{M})/c^2]\bm{\hat{j}}.
\end{equation}
There is no angular electromagnetic momentum in this case due to the uniformity
of the electric field, but when you Lorentz-transform the angular momentum
four-tensor, $L^{\mu,\nu}$, containing the linear momentum times $ct$ in the
$L^{2,4}$ slot to the $S'$ frame, you get the following angular momentum in the
positive $z$ direction,
\begin{equation}
\label{L of loop}
L_{z'} = \gamma vtE\mathcal{M}/c^2 = \mathcal{M}B't.
\end{equation}
Note that this angular momentum is increasing in the positive $z$ direction.
Meanwhile the angular momentum due to the Lorentz forces is increasing in the
negative $z$ direction by the same amount. That is,
\begin{equation}
\label{em torque}
\tau_{z'} = \frac{dL_{z'}}{dt} - \mathcal{M}B' = 0,
\end{equation}
where Eqs. (\ref{L of loop}) and (\ref{torque on loop}) have been used. There
is no torque on the loop in either frame.

An interesting example of hidden momentum discussed by Vaidman is a point charge
at the center of a current carrying toroidal coil. The coil lies in the $x$-$y$
plane with its center at the origin. Also at the origin is a point charge.
Vaidman calculates the electromagnetic momentum of this combination as
approximately $-\epsilon_o\mu_o N\bm{\mathcal{M}}\times\bm{E}$, where $N$ is
the number of
turns in the coil, $\mathcal{M}$ is the magnetic moment of a turn, and $E$ is
the electric field at the coil due to the point charge. For a current that
creates a magnetic field in the positive $\phi$ direction around the coil and a
positive point charge, this momentum is in the positive $z$ direction.

You can assemble the charge-coil system loop-by-loop using the results given
above for the Shockley-James model. As each loop is brought toward the charge
in turn, an impulse of $(1/2)\epsilon_o\mu_o\bm{\mathcal{M}}\times\bm{E}$ is
imparted to the charge and to the loop. For $N$ loops this is just the opposite
of the electromagnetic momentum of the coil.

I would be remiss if I didn't  acknowledge that there are other ways to put the
system together that don't appear to lead to a balance of mechanical and
electromagnetic
momentum. For example, you can grow a charge in the center of the coil from
zero to $q$ (like charging one side of a capacitor). The opposing mechanical
momentum received by the coil is half that of the electromagnetic momentum. Or,
you can bring the charge in from a far distance along the axis of the coil
towards its center. The combined mechanical momentum supplied to the coil from
the Lorentz forces and lost to the charge due to the time rate of change of the
vector potential at the position of the moving charge yields the same result
\cite{note}. It is clear that in the loop-by-loop assembly the changing magnetic
field of each loop in turn exerts a force on the charge, a force that does not
appear in my calculations of the other two methods of assembly. Possibly this
can be resolved, but none of these methods of assembly is consistent with the
presence of hidden momentum.

Recently, Boyer \cite{Boyer5} has performed calculations for a model of a
charge-magnet system in which he claims to have found a new form of hidden
momentum.  (He prefers the term ``internal momentum'' to ``hidden momentum'').
In Boyer's
model the magnet consists of a negative charge of $-Ne$, where $N$ is a
positive integer, at the center of a circular frictionless track around which
$N$ charges of $+e$ may orbit. In the vicinity of this magnet is a positive
charge distant enough (and/or the magnet small enough) that its electric field
at the magnet is essentially uniform. He includes calculations for both
non-interacting and interacting magnet charges.

In the non-interacting version he finds the mechanical hidden momentum of
Coleman and Van Vleck. In the interacting version he has two interacting
electric charges circling the track
and finds a hidden momentum consisting of both mechanical and electromagnetic
contributions. However, the central argument regarding the ignoring of Lorentz
forces when the magnet is placed in an electric field applies whether or not
the charges in the magnet are interacting.

Mansuripur \cite{Mansuripur} proposed a paradox in which there is a charge in
the vicinity of a current loop. An observer stationary with respect to these
objects sees no interaction between them (with the possible exception of an
induced electrical polarization of the loop). An observer moving with respect to
the charge and loop will, in general, see an electric dipole resulting from the
Lorentz transformation of the magnetic dipole of the loop (but not a dipole due
to a separation of charge, as I discuss below). Hence, this observer
should see a torque acting on the loop due to the electric field of the charge
acting on the electric dipole. One observer sees a torque on the loop and
another does not. Mansuripur argues that this invalidates the Lorentz force law
and proposes the force law of Einstein and Laub \cite{Einstein} is the correct
one instead. To save the Lorentz force, it was proposed that the torque due to
the charge-dipole interaction was countered, in the moving observer's frame,
by a torque due to hidden momentum \cite{Vanzella,Barnett,Saldanha,Khorrami,DJ}.

However, I have shown \cite{redfern1} that there is no need for hidden momentum
to solve this paradox. There is both linear and angular momentum in the
electromagnetic fields of a charge-magnet system \cite{Furry}. When seen in the
moving frame, these fields contain a torque that counters the one resulting
from charge-dipole interaction. Both these torques are actually time rate of
change of angular momentum in the electromagnetic fields themselves. (In my
reference \cite{redfern1} I referred to this torque as ``fictitious''. This was
a poor choice of words. The point is it is electromagnetic torque, not
mechanical torque.)

Saldanha and Filho \cite{Saldanha2} recently placed a paper on the arXiv
preprint server regarding the role of hidden momentum in physical media. Their
starting point was the non-relativistic Lagrangian for a particle with both an
electric dipole and magnetic dipole moving through a magnetic and electric
field. The Lagrangian is
\begin{equation}
\label{SF L}
\mathcal{L} = \frac{1}{2}mv^2 + \bm{E}\cdot\bm{p} + \bm{B}\cdot\bm{\mathcal{M}}.
\end{equation}
Here, $m$ is the mass of the particle with speed $v$, $\bm{E}$ and $\bm{B}$ are
the electric and magnetic fields, and $\bm{p}$ and $\bm{\mathcal{M}}$ are the
electric and magnetic dipole moments, respectively, as seen in the lab frame. If
$\bm{p}_o$ and $\bm{\mathcal{M}}_o$ are the dipole moments in the particle's
rest frame, then $\bm{p} = \bm{p}_o +
\epsilon_o\mu_o\bm{v}\times\bm{\mathcal{M}}$ and $\bm{\mathcal{M}} =
\bm{\mathcal{M}}_o - \bm{v}\times\bm{p}_o$, where $\bm{v}$ is the particle's
velocity. (These relationships are for $v << c$, such that the Lorentz factor
$\gamma$ is taken to be one.)

To get the canonical (total) momentum of the particle they take the velocity
gradient of the Lagrangian.
\begin{equation}
\label{grad L}
\bm{\nabla}_{\bm{v}} \mathcal{L} = m\bm{v} - \bm{p}_o\times\bm{B} + \epsilon_o\mu_o\bm{\mathcal{M}}_o\times\bm{E}.
\end{equation}
They attribute the last term on the right-hand side of the above equation to
hidden momentum. However, this interpretation is erroneous, once again, due to
ignoring the Lorentz forces that arise upon applying electromagnetic fields to
magnets. Their argument for hidden momentum parallels that of Coleman and Van
Vleck \cite{Coleman}, where work is done on orbiting non-interacting charged
particles by the fields, resulting in greater speed on one side of the orbit
than the other. The argument is that even though linear charge density remains
the same around the orbit, the linear momentum density does not due to the
relativistic increase in mass. I have shown above that this view is not viable.

Saldanha and Filho examine four different types of media for the presence of
hidden momentum, assuming there could be magnetic charges. No hidden momentum
is found in media where the dipoles consist of electric and magnetic charges
separated by a small distance as there are no orbiting charges on which work can
be done. The media where they find hidden momentum are the three others where
there are orbiting charges forming the dipoles -- one with all dipoles due to
orbiting charges and two others missing either orbiting electric charges or
orbiting magnetic charges. In particular, they claim orbiting magnetic
charges form electric dipoles that contain hidden momentum. They apply the
supposed hidden momentum in these media to the question of Minkowski and Abraham
momenta.

Consider magnetic charges forming a circuit and creating the Amperian version of
an electric dipole at the origin of a Cartesian coordinate system. To assemble
a model analogous to that of Shockley and James \cite{Shock}, bring magnetic
charges from a large distance toward the magnetic circuit in the plane of the
circuit. As before, have the positive (north pole) charge move along the $x$
axis in the positive direction and the negative (south pole) charge move along
the $x$ axis in the negative direction.

Due to the changes of sign in Maxwell's equations when interchanging the roles
of the electric and magnetic quantities, the counterclockwise magnetic circuit
produces an electric dipole in the negative $z$ direction. Also the Lorentz
force on the magnetic charges moving with velocity $\bm{v}$ will be
$-(q_m/c^2)\bm{v}\times\bm{E}$, where $\bm{E}$ is the electric field (in the
positive $z$ direction at the location of the magnetic charges) and $q_m$ is the
magnetic charge of a north pole. The result of
these changes means the force on both magnetic charges will be in the positive
$y$ direction and will equal to $vBp/r$ in magnitude, where $p$ is the magnitude
of the electric dipole moment. Integrating over time to find the impulse, as was
done for the model of Shockley and James, yields a total impulse of
$\bm{B}\times\bm{p}$ acting on the charges in the positive $y$ direction.

Performing an analogous calculation for the impulse on the disks as was done in
the SJ model, the surface integral of the ``displacement current'' is
$-\pi\mu_o q_m v^2sin^2\phi/(2\pi r^3) $, resulting in an electric field on the
rims of the disks of $-\mu_o q_m vasin\phi\bm{\hat{k}}/4\pi r^3$. The electric
field acting on the magnetic current results in an impulse on the disk
equal to that on the charges and in the same direction. (Here, $\phi$ is the
same azimuthal angle as in the SJ model and $a$ is the radius of the disks.)

So, the total impulse due to applying a magnetic field to the magnetic
charge-``Amperian'' electric dipole system is $2\bm{B}\times\bm{p}$. The
negative of this is the electromagnetic momentum. This is easily shown by
starting with Eq. (\ref{em momentum}) for the magnetic charge-Amperian electric
dipole system and manipulating it algebraically until it is identical to that
of the SJ system. There is therefore no hidden momentum necessary for momentum
conservation for an Amperian electric dipole in a magnetic field.

Filho and Saldanha \cite{Filho} recently performed a quantum calculation to
find the hidden momentum in a hydrogen atom using degenerate perturbation
theory. As in classical calculations, they consider the atom to be at rest in
the electric field, ignoring what happens when the electric field rises from
zero to its final value. It would appear that time-dependent perturbation theory
should be used to give a complete picture. There is no reason a quantum
calculation should show hidden momentum when correct classical calculations do
not as hidden momentum is a non-quantum conjecture.

Finally, Spavieri \cite{Spavieri} found a role for hidden momentum in the
spin-orbit effect in a hydrogen atom. Much of his discussion involves the
interaction between the induced electric dipole $\bm{p} =
\bm{v}\times\bm{\mathcal{M}}$ on the electron due to its magnetic
dipole being in motion around the nucleus with speed $v$ and the electric field
of the nucleus. Spavieri finds the
electromagnetic interaction energy $U$ between the induced electric dipole and
the electric field $\bm{E}$ of the nucleus by performing a volume integral as
follows.
\begin{equation}
\label{U}
U = \epsilon_o \int(\bm{E}_p\cdot\bm{E})dV = -\bm{p}\cdot\bm{E},
\end{equation}
where $\bm{E}_p$ is the electric field due to the induced dipole. There is a
problem with this result.

The magnetic field of a magnetic dipole with magnetic moment $\bm{\mathcal{M}}$
at the origin of the coordinate system in its rest (primed) frame is
\begin{equation}
\label{B of dipole}
\bm{B}' = \frac{\mu_o}{4\pi}\left[\frac{3(\bm{\mathcal{M}}'\cdot\bm{r}')\bm{r}'}{r'^5} - \frac{\bm{\mathcal{M}}'}{r'^3}\right] - \frac{2\mu_o}{3}\bm{\mathcal{M}}'\delta^3(\bm{r}'),
\end{equation}
where $\delta^3(\bm{r}')$ is the Dirac delta function.
Transforming this to the moving (unprimed) frame as seen in the coordinate
system at rest with the nucleus (see \cite{Rindler} for the technique) you find
\begin{eqnarray}
\label{transformed B}
\bm{B} &=& \frac{\mu_o}{4\pi\gamma^2}\left[\frac{3(\bm{\mathcal{M}}\cdot\bm{r})\bm{r}}{\gamma^2 r^5[1-(v^2/c^2)sin^2\alpha]^{5/2}} \right. \\ \nonumber
&-& \left. \frac{\bm{\mathcal{M}}}{r^3[1-(v^2/c^2)sin^2\alpha]^{3/2}}\right] - \frac{2\mu_o}{3}\bm{\mathcal{M}}\delta^3(\bm{r}),
\end{eqnarray}
where $\alpha$ is the (instantaneous) angle between $\bm{r}$ and $\bm{v}$. The
corresponding electric field is
\begin{equation}
\label{transformed E}
\bm{E}_p = -\bm{v}\times\bm{B}.
\end{equation}

Now the volume integral of the field of an actual dipole due to a separation of
charge will indeed equal the dipole itself due to the delta function; however,
it is clear that the electric field $\bm{E}_p$ is not that of an actual dipole.
For one thing there is no electric field parallel to the direction of motion of
the magnetic dipole. In the slow motion case where $v << c$, the field can be
written as
\begin{equation}
\label{transformed E slow}
\bm{E}_p = \bm{E}_{dp} - \frac{3\bm{p}}{4\pi\epsilon_o r^3}sin^2\theta - \frac{3\bm{\hat{m}}\times\bm{p}}{4\pi\epsilon_o r^3}sin\theta cos\theta sin\phi.
\end{equation}
Here $\bm{E}_{dp}$ is the electric field of a dipole $\bm{p}$ that consists of
an actual separation of charge. The angles are those of a spherical system of
coordinates, where the velocity is in the positive $z$ direction. (Thus $\theta$
is the same as $\alpha$ in the earlier equations.) The third term on the right
hand side of the above equation cancels the field in the $z$ direction of the
first term such that there is no field parallel to $z$. The problem in taking
the volume integral of $\bm{E}_p$ is due to the second term on the right
hand side. It results in a term proportional to $ln(1/r)$, which is undefined
at the limits of the integral. (The third term integrates to zero due to
$sin\phi$.)

Of course, there can't be an actual charge separation on an
electron, and the Lorentz transformation of the fields is consistent with that.
(I have performed a calculation that indicates the apparent dipole on a current
loop is due to a relativity-of-simultaneity effect rather than an actual charge
separation \cite{redfern4}).
Since the rest of the argument of Spavieri depends on this calculation, his
results are brought into question. 

\section{Discussion and conclusion}

This paper examines models where a magnet is embedded in an electric field. The
model due to Shockley and James \cite{Shock}, with its electrically charged
rotating disks in the vicinity of two equal and opposite external charges,
acquires mechanical momentum when the charges are moved in from a distance,
unless the charges and magnet are held stationary by an external agent. This
mechanical momentum is equal and opposite to the electromagnetic momentum it
acquires at the same time.

No hidden momentum appears in this model and no
force is applied to the magnet when it is demagnetized, which is the paradox
Shockley and James thought they had discovered. Rather, the so-called paradox
is simply due to analyzing this model in a reference frame that is moving with
respect to the frame in which it was assembled (by either applying an electric
field to the magnet or forming the magnet in an electric field). If the model is
held stationary, then the ``hidden momentum'' actually resides in the external
agent that held the model at rest.

Coleman and Van Vleck examined the hidden momentum proposal of Shockley and
James with a calculation involving a point charge and a model-free Amperian
magnet using the Lagrangian of Darwin \cite{Coleman}. Their work appeared to
confirm a force on the magnet when it is demagnetized. A calculation based on
a specific model alluded to by Coleman and Van Vleck shows their result only
obtains for a very contrived model that already has momentum. The force they
claim is experienced by the magnet when it is demagnetized is actually the loss
of momentum by the electromagnetic field.

The calculation of Coleman and Van Vleck assumed the electric charges that
produced the magnetism were non-interacting, as did Babson \textit{et al} in
a model they examined \cite{Babson}. If formed by applying an electric field to
non-interacting particles constituting a current, the resulting particle motion
is not the same as that of their models and no hidden momentum results.

The incompressible, frictionless fluid magnet model of Babson \textit{et al}
contains a relativistic momentum in the fluid due to pressure, which is
balanced by the relativistic momentum in the
tube containing the fluid due to it being subject to stress. This momentum
might be viewed as the ``internal momentum'' proposed by Boyer \cite{Boyer5}.
It disappears with no effect on the motion of the magnet when the electric field
is removed. Thus, this is not the type of hidden momentum envisioned by
Shockley and James and by Coleman and Van Vleck, since their momentum would
create an impulse on the magnet when its magnetism goes to zero.

The total momentum of a magnetic dipole
created in a uniform electric field is shown to be zero without the need for
hidden momentum. You must consider the effect of the creation of
the dipole on the charges responsible for the uniform Coulomb field, whether
that of the finite system used here or of infinite parallel plates
\cite{Mansuripur2}. Taking the moving frame of this model as its rest frame for
analysis is misleading, as it is for the other models. There is an
electromagnetic momentum given by the sum of Eqs. (\ref{Pem inside}) and
(\ref{Pem outside result}), but due to observation in the new rest frame there
is no observed mechanical momentum. This is not the case in the original rest
frame where the magnet was formed in an electric field. Hence whether an
electric field is applied to a magnet or a magnet is created in an electric
field, no hidden momentum appears.

Finally, some previous results in the literature were examined. In the dispute
over whether or not a neutron experiences a force in an electric field
\cite{Boyer6,APV,AC}, I found that if the neutron is modeled as a current loop,
it does not contain hidden momentum, since a current loop contains no hidden
momentum. The paradox explored by Vaidman \cite{Vaidman} where a current loop
is moving in an electric field, like that of Mansuripur
\cite{Mansuripur,redfern1} is resolved by transforming linear electromagnetic
momentum of a current loop in an electric field in a frame at rest with the
loop to a frame in which the loop is
moving. The transformation produces an electromagnetic torque that cancels the
anomalous torque in the moving frame. Also shown is that an Amperian electric
dipole consisting of circulating magnetic charges, like its magnetic
counterpart, holds no hidden momentum. Lastly, a calculation by Spavieri
\cite{Spavieri} showing a connection between hidden momentum and spin-orbit
splitting in hydrogen-like atoms fails due to the electric field of a moving
magnetic dipole not being identical to that of an electric dipole with a charge
separation.

The center-of-energy theorem is often invoked to support the idea of hidden
momentum \cite{Griff}. In this view hidden momentum is necessary to satisfy the
theorem. However, it should be clear from the discussion in this paper that
center-of-energy arguments erroneously assume the charge-magnet systems are at
rest in their original frames of reference. This ignores the Lorentz forces
necessarily involved when either an electric field is applied to an Amperian
magnet or a magnet is formed in an electric field. A charge-magnet
system can be assembled and gain both electromagnetic field and mechanical
momentum (equal and opposite). So, it is possible for an observer to adjust her
velocity so that the system is stationary in her reference frame. In her frame
the system contains field momentum but no mechanical momentum.

Of course, the presence of a net force on a charge-magnetic dipole system during
its assembly with no equal and opposite force violates the usual view of
Newton's third law of motion. However, it is clear from the calculations given
here that the four-force acting on the system is zero if you include the
electromagnetic momentum in the space component along with the mechanical
momentum in the momentum four-vector as the system is assembled. This renders
the spatial component zero. The time derivative of this is the relativistic
three-force, which is also zero.

I have calculated other charge-magnet systems that are claimed to contain
hidden momentum \cite{redfern2} and have found no hidden momentum when
the assembly of the systems is considered. Although these calculations concern
specific systems, if hidden momentum exists for a magnet in an electric
field, it should be evident in these calculations. I conclude that hidden
momentum does not exist in these systems and that such systems at rest with
respect to an observer can contain electromagnetic momentum.

\hfill\break

\noindent This work was done by the author without external financial support.

\bibliographystyle{epj}
\bibliography{d160263}

\begin{thebibliography}{42}
\bibitem{note}
Added in proof: It turns out that the effect of the vector potential is
  canceled by that of the scalar potential, but the conclusion concerning
  hidden momentum still holds.
\end{thebibliography}

\end{document}